\documentclass[aps,a4paper,floatfix,pre,showpacs,balancelastpage,superscriptaddress,twocolumn]{revtex4-1}

\usepackage{graphicx}
\usepackage[dvipsnames]{pstricks}
\usepackage{amsmath,amsfonts,amssymb}
\newcommand{\REM}[1]{}
\usepackage[%
  unicode=true,%
  colorlinks=true,%
  linkcolor=blue,anchorcolor=blue,%
  breaklinks=true,%
  citecolor=blue,filecolor=cyan,%
  menucolor=blue,%
  urlcolor=blue]{hyperref}

\begin{document}
\title{
Classical lattice spin models
involving singular interactions isotropic in spin space}

\author{Hassan Chamati}
\affiliation{Institute of Solid State Physics, Bulgarian Academy of Sciences,
72 Tzarigradsko Chauss\'ee, 1784 Sofia, Bulgaria\\
chamati@bas.bg}

\author{Silvano Romano}
\affiliation{Physics dept., the University,
via A. Bassi 6, 27100 Pavia, Italy \\
silvano.romano@pv.infn.it}

\begin{abstract}
We address here a few classical lattice--spin models, involving
$n-$component unit vectors ($n=2,3$), associated with a
$D-$dimensional lattice $\mathbb{Z}^D,~D=1,2$, and interacting via a
pair potential restricted to nearest neighbours and being isotropic in
spin space,  i.e. defined by a function of the scalar product between
the interacting spins. When  the potential involves a continuous
function of the scalar product, the Mermin--Wagner theorem and its
generalizations exclude orientational order at all finite temperatures
in the thermodynamic limit, and exclude phase transitions at finite
temperatures when $D=1$; on the other hand, we have considered here
some comparatively simple functions of the scalar product which are
bounded from below, diverge to $+\infty$ for certain mutual
orientations, and  are continuous almost everywhere with integrable
singularities.
Exact solutions are presented for $D=1$, showing absence of phase
transitions and absence of  orientational order at all finite
temperatures in the thermodynamic limit; for $D=2$, 
and in the absence of more stringent mathematical results,
extensive
simulations carried out on some of them point to the absence of
orientational order at all finite temperatures, and suggest the
existence of a Berezinski\v\i-Kosterlitz-Thouless transition.

\end{abstract}
\date{\today}
\maketitle

\section{Introduction} \label{intro}
The study of lattice spin models, both classical
(on which we shall be concentrating here) and quantum, is an important
chapter of Statistical Mechanics, where a number of mathematical 
results have been obtained, entailing absence or existence, and 
sometimes type, of phase transitions at finite temperatures, depending 
on lattice dimension, number of spin components, range and symmetry of 
the interaction.

The Mermin-Wagner theorem was  first  proven nearly 50 years ago
in a quantum setting, for the isotropic spin Heisenberg model
with finite--range exchange interactions \cite{rMW00,*rMW001},
and later extended by various Authors in a number of directions, e.g. 
to the classical setting, to other functions of the scalar product, or  
to longer--ranged interactions \cite{rMW01,*rMW02,rMW03,rMW04,rMW05};
see also a subsequent Review in Ref. \cite{rGN}.

In the classical case, the Mermin--Wagner theorem and its 
generalizations \cite{rSinai,rGeor,rMWlate,rMWarXiv} hold for lattice--spin 
models, consisting of $n-$component unit vectors ($n \ge 2$), 
associated with a $D-$dimensional lattice $\Lambda_D$ ($D=1,2$ and 
typically $\Lambda_D=\mathbb{Z}^D$),
and interacting via pair potentials which are isotropic in spin space, 
and usually translationally invariant
(on the other hand, mathematical results have also been
obtained which do not need any translational invariance
\cite{ntir01,ntir02,ntir03,*ntir04,ntir05,ntir06,ntir07});  the
distance dependence is usually taken to be suitably short-ranged.
Their orientational dependences are defined by some functions 
of the scalar product between interacting spin pairs:
the earlier mathematical results were obtained for  rather
smooth functions (simple polynomials), and conditions
were later gradually relaxed, i.e. to the milder request
of continuity, and, in some cases, even to less regular functions
\cite{rMWlate,rMWarXiv}.
 
More explicitly, continuity is required
in Refs. \cite{rMWlate,rMWarXiv}, and some singularities
are also allowed for in Ref. \cite{rMWlate};  we are restricting our
present discussion to finite--range (actually, nearest--neighbour)
interactions, and notice that mathematical results 
are known for long--range interactions as well (see, {\it e.g.}, Refs.
\cite{rGeor,rMWlate,rMWarXiv}, and others quoted therein).

To fix notation and ideas, let
$\mathbf{w}_j=(w_j^1,w_j^2,\cdots,w_j^n)$ denotes the $n$--component
unit vector (spin) associated with the $j-$th lattice site, with dimensionless
coordinate vector $\mathbf{x}_j \in \mathbb{Z}^D$;
two--component spins are parameterized by usual polar angles $\varphi_j$,
and three--component spins are  parameterized by  usual spherical
angles $\left( \theta_j,~\phi_j \right)$.
Here and in the following the interaction will be restricted to nearest
neighbours and defined by
\begin{equation}
\Phi\equiv\Phi(\tau)= \epsilon F(\tau),~\tau\equiv\tau_{jk}=
\mathbf{w}_j \cdot \mathbf{w}_{k},
\label{eq01}
\end{equation}
where $\epsilon >0$ denotes a positive quantity setting energy and temperature
scales (i.e. $T= k_B \mathcal{T}_K/\epsilon$, where 
$\mathcal{T}_K$ denotes the temperature in
degrees Kelvin), and to be scaled away from the following formulae.
For $2-$component spins, it will prove notationally convenient
to define
\begin{equation}
\tau=\cos \Delta, \quad ~\Delta\equiv\Delta_{jk} = \varphi_j-\varphi_k.
\label{EQDEF2}
\end{equation}
When $F(\tau)$ is a continuous function of its argument,
the above theorems entail absence of orientational order in the
thermodynamic limit at all finite temperatures
\cite{rMWlate}; 
when $D=n=2$, and under additional conditions, a 
Berezinski\v\i-Kosterlitz-Thouless (BKT), or, in more general
terms, a BKT-like transition can be proven to exist
\cite{rBKTrig,revBKT1,revBKT2,revBKT3,genBKT,BKTbook,vanenter2002,vanenter2005};
the term ``BKT-like'' is used here to indicate a transition to a disordered
low-temperature phase possessing slowly decaying correlations resulting
in infinite susceptibility; in thermodynamic terms, the transition may be
of infinite order (as in the more common, originally studied BKT case
\cite{rBKTrig,revBKT1,revBKT2,revBKT3,genBKT,BKTbook};
  it was also later proven
\cite{vanenter2002,vanenter2005} that it can turn first-order under
certain conditions.

Cases where $F(\tau)$  possesses some singularity
have been studied far less extensively (see also below). In fact
one can envisage a multitude of singular interactions: models involving a 
finite number of jump discontinuities, as in  sign 
or step models, are discussed in Appendix \ref{AppA}; another family, also
discussed there, involves constrained models, where whole 
regions of configuration space are excluded.
We have chosen to start our investigation, 
so to speak, somewhere in between these two cases, from
functional forms containing slowly divergent terms
which do not disturb thermodynamics, {\it i.e.} from functional
forms being bounded from below, continuous almost everywhere, 
slowly diverging to $+\infty$ for one (or a few) mutual orientations,
and possessing integrable singularities. Thus the present paper addresses a 
few models whose functional forms are defined by
\begin{subequations}
\label{modelsdef}
\begin{eqnarray}
V(\tau)& =& -\ln(1+\tau), \quad n=2,
\label{Vmodel}
\\[0.25cm]
W(\tau)& =& -\ln(1+\tau), \quad n=3,
\label{Wmodel}
\\[0.25cm]
X(\tau)& =& -\ln(|\tau|), \quad n=3.
\label{Xmodel}
\end{eqnarray}
\end{subequations}
In due course, comparisons will also be made with their 
extensively studied counterparts defined by
\begin{subequations}
\label{regall}
\begin{eqnarray}
F(\tau) & = & -\tau, \quad\qquad ~n=2,
\label{reg01}
\\[0.25cm]
F(\tau) & = & - \tau, \quad\qquad n=3,
\label{reg02}
\\[0.25cm]
F(\tau) & = & -P_2(\tau), \quad n=3,
\label{reg03}
\end{eqnarray}
\end{subequations}
respectively, and simply referred to as ``regular counterparts''.

Some models bearing similarities to ours [Eq. \eqref{Wmodel}]  have 
been investigated previously in the literature
\cite{rsing1,rsing2,rsing3,rsing4,rsing5,*rsing6}.
More recent studies showed that such classical models are effective
models obtained via mappings
from quantum-mechanical treatments \cite{rsing2,rsing3,rsing4}.
The above singular models [Eqs. (\ref{modelsdef})], as well as 
some generalizations and linear combinations of them,
can be solved exactly when $D=1$, allowing one
to obtain  thermodynamic and structural quantities in closed form;
these are worked out in Appendix \ref{AppA}, where
other singular models, such as step or sign model and constrained
ones are addressed as well. The three models in Eqs. (\ref{modelsdef})
are studied by extensive Monte Carlo (MC) simulation for $D=2$
so as to explore the thermodynamic behavior of these models, on the one
hand, and to unveil potential effects of the singularities in
comparison with their regular counterparts, on the other hand.

The rest of the paper is organized as follows: in Sec. \ref{models}
we further discuss the singular models;
our simulation methodology for $D=2$ is discussed in
Section \ref{comptaspect} along with with brief details on the finite-size
approach we employ for the analysis of the simulation data. In Sec.
\ref{results} we present the simulation results and finite-size
scaling analysis used to extract the critical behavior for
the models under consideration. We conclude the paper with
Sec. \ref{summary} where we summarize our results.

\section{Remarks on the potential models }\label{models}
Both $V$ and $W$ attain their minimum at $\tau_{min}=1$,
and slowly diverge to $+\infty$ as $\tau \rightarrow -1$;
$X(\tau)$ attains its minima at $\tau_{min}=\pm 1$,
and slowly diverges to $+\infty$ as $\tau \rightarrow 0$;
the above functions are bounded from below, continuous almost everywhere,
and possess integrable singularities;
in these cases, an interaction diverging to $+\infty$ is still 
compatible with the thermodynamics
and, by its very functional form, it can  be expected to enforce  
some strengthening of short-range correlations. On the other hand,
changing the sign in front of the ``$\ln$'' from ``$-$'' to
``$+$'' in (any of) Eqs. (\ref{modelsdef}) would produce a rather
dramatic effect, i.e. it would cause a divergence to $-\infty$ for
some mutual orientations, and hence make the modified model not well
defined at low temperatures \cite{rsing1}.

\begin{figure}[h!]
\resizebox{\columnwidth}{!}{\includegraphics{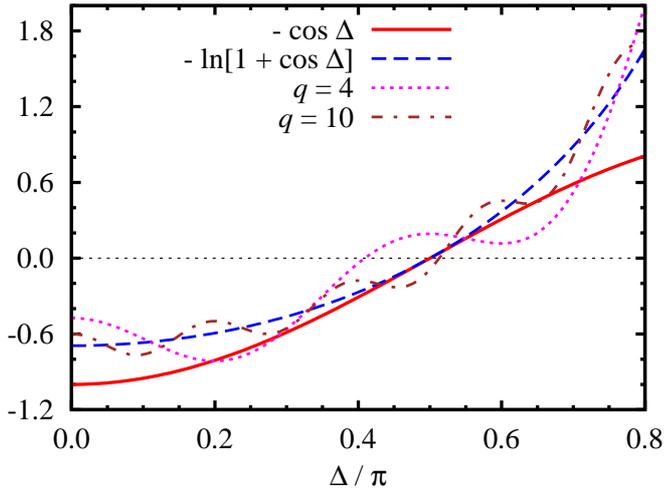}}
\caption{(Color online) Comparison beteween the singular model $V$, its regular counterpart,
and some truncated expansions [Eqs. (\ref{Vmodel}), (\ref{reg01}),
(\ref{vsexp})], as functions 
of the angle $\Delta$ between the two spins.
Meaning of symbols: red
continuous line: regular counterpart; blue dashed line:
model $V$; magenta dotted line: $\mathcal{V}_4$;
brown dash-dotted line: $\mathcal{V}_{10}$.}
\label{modelv}
\end{figure}

\begin{figure}[h!]
\resizebox{\columnwidth}{!}{\includegraphics{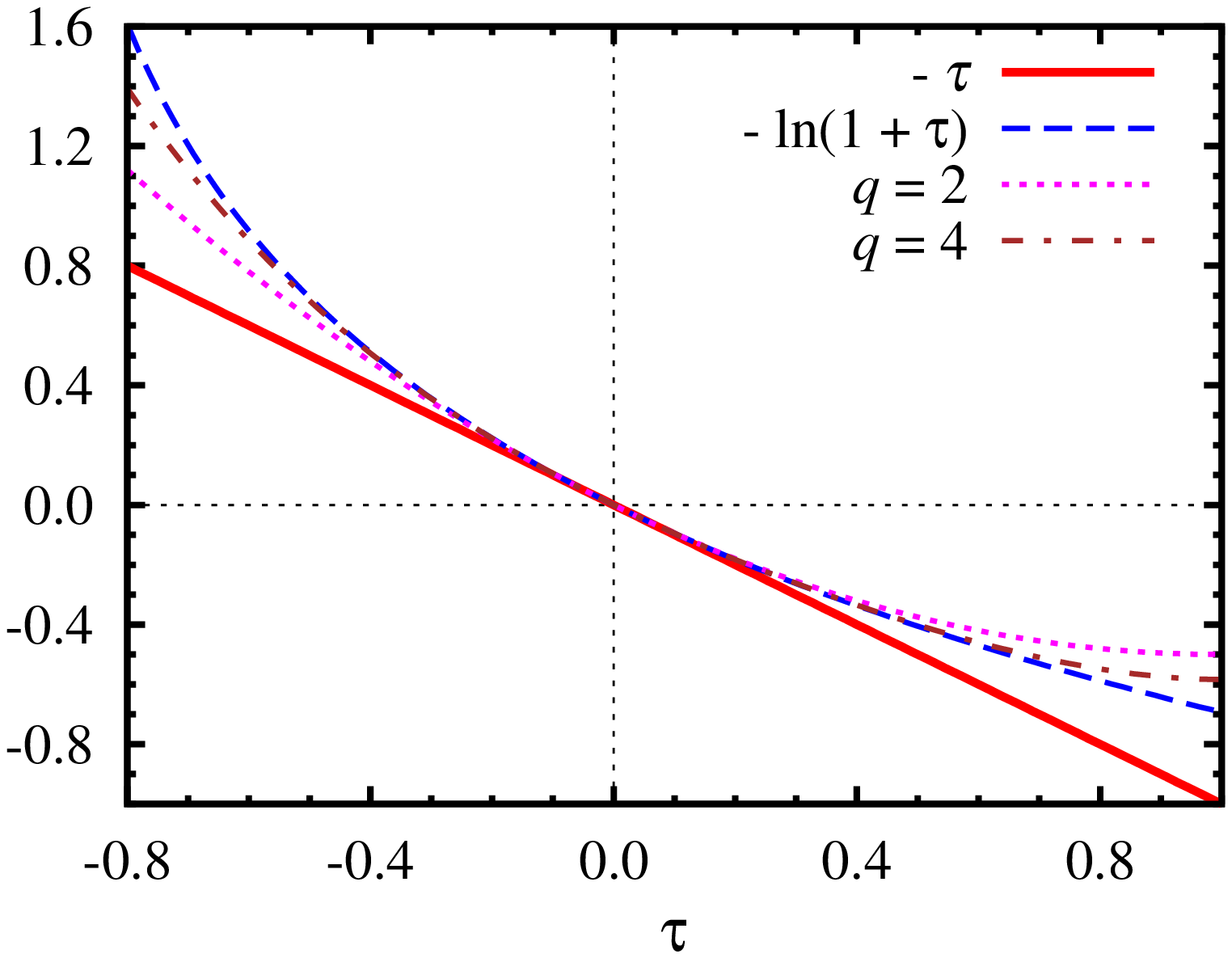}}
\caption{(Color online) Comparison beteween the singular model $W$, its regular counterpart,
and some truncated expansions [Eqs. (\ref{Wmodel}), (\ref{reg02}),
(\ref{wsexp})], as functions of the scalar product
$\tau$ between the two spins. Meaning of symbols: red
continuous line: regular counterpart; blue dashed line:
model $W$; magenta dotted line: $\mathcal{W}_2$;
brown dassh-dotted line: $\mathcal{W}_4$.}
\label{modelw}
\end{figure}

\begin{figure}[h!]
\resizebox{\columnwidth}{!}{\includegraphics{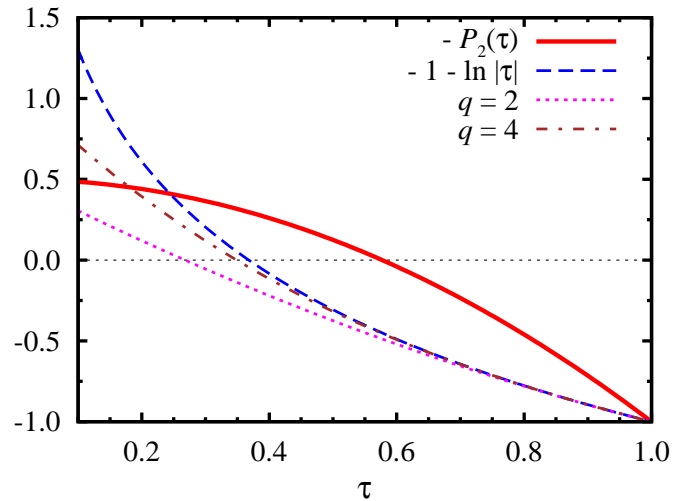}}
\caption{(Color online) Comparison beteween the singular model $X$, its regular counterpart,
and some truncated expansions [Eqs. (\ref{Xmodel}), (\ref{reg03}),
(\ref{xsexp})], as functions of the scalar product
$\tau$ between the two spins. Meaning of symbols: red
continuous line: regular counterpart; blue dashed line:
model $X$; magenta dotted line: $\mathcal{X}_2$;
brown dash-dotted line: $\mathcal{X}_4$.
Notice that the quantity $-1$ has been added to $X$ 
as well as to the two  truncated expansions, in order to ease
comparison.
}
\label{modelx}
\end{figure}

Series expansions of Eqs. 
(\ref{modelsdef}) can be written down, {\it i.e.}
\begin{subequations}
\label{modelsexp}
\begin{eqnarray}
V(\tau) &  = & \ln(2) + \lim_{q \rightarrow \infty} \mathcal{V}_q,
\nonumber
\\
\mathcal{V}_q & = &2 \sum_{l=1}^q \frac{(-1)^l}{l} \cos(l \Delta_{jk}),~
0 \le \Delta_{jk} < \pi; 
\label{vsexp}
\\
W\tau) & = & \lim_{q \rightarrow \infty} \mathcal{W}_q,
\nonumber
\\
\mathcal{W}_q &  = & 
\sum_{l=1}^q \frac{(-1)^l}{l} \tau^l,~
-1 < \tau \le 1;
\label{wsexp}
\\
X(\tau) & = &  \lim_{q \rightarrow \infty} \mathcal{X}_q,
\nonumber
\\
\mathcal{X}_q
& = & \sum_{l=1}^q \frac{1}{l} (1- |\tau|)^l,~0 < |\tau| \le 1;
\label{xsexp}
\end{eqnarray}
\end{subequations}
each $\mathcal{X}_q$ is a polynomial in 
$|\tau|$, where the coefficient in front of $|\tau|^l$ bears the sign 
$(-1)^l$; in other words sign alternation is a common feature of the three
above expansions.
Any of the above truncated expansions [Eqs. (\ref{modelsexp})]
is a continuous function of $\tau$
which, by the Mermin-Wagner theorem and its generalizations 
\cite{rMWlate,rMWarXiv},
produces orientational disorder at all finite temperatures;
let us now consider a generalization of $\mathcal{V}_q$, {\em i.e.}
\begin{equation}
\mathcal{F}_q = \sum_{l=1}^q c_l \cos(l \Delta),
\label{GENPOL}
\end{equation}
where $c_l$ denote arbitrary real coefficients; the Mermin-Wagner
theorem can be applied here as well; moreover, for a
general ferromagnetic interaction (where all the coefficients
$c_l$ are $\le 0$), one can
{\em  prove} BKT behavior, based on its
existence for Eq. (\ref{reg01}) \cite{rBKTrig} and on correlation inequalities,
and also obtain a rigorous lower bound on the BKT transition temperature
(see Ref. \cite{genBKT} and others quoted therein); unfortunately,
the alternating signs in $\mathcal{V}_q$ {\it prevent}
us from using this approach in general. 
Let us also mention in passing a simple specific case of Eq.
(\ref{GENPOL}), defined by
\begin{equation}
\mathcal{G}_2 = c_1 \cos \Delta + c_2 \cos(2 \Delta),~c_1 < 0,
\label{fc1c2}
\end{equation}
where $c_2$ can both be negative or sweep a suitable range of
positive values; the model was
studied by various Authors in the Literature
(see Refs. \cite{PhysRevB.64.092407,PRL106067202}
and others quoted therein), also 
in the equivalent version \cite{pre72031711,pre80011707}
(recall Appendix \ref{AppB})
\begin{equation}
\mathcal{G}_4
= c_2 \cos(2 \Delta)  + c_4 \cos(4 \Delta),~c_2 < 0;
\label{fc2c4}
\end{equation}
simulation or spinwave evidence of BKT behavior was obtained in various cases, 
and estimates of the BKT transition temperature obtained for  
cases where the above mathematical treatment applies 
\cite{PhysRevB.64.092407,pre72031711}
were later shown to agree with the named lower bound \cite{genBKT}.
It proves convenient to compare each singular interaction
potential [Eqs. \eqref{modelsdef}] with its regular counterpart [Eqs.
\eqref{regall}], and with some
truncated expansion
[Eqs. (\ref{modelsexp})]; this is done in FIGs. \ref{modelv},
\ref{modelw} and \ref{modelx}. These are found to exhibit a common feature:
on the one hand, the singular interactions diverge rather slowly for
appropriate mutual orientations; on the other hand, in a broad
minimum-energy region, the growth of the singular interaction
energy as $\tau$ moves away from the corresponding $\tau_{min}$
is recognizably {\em slower} than for its regular counterpart,
and then it becomes faster and faster  outside this region;
the changeover takes place about 
$\tau = 0$ ($V$ and $W$ model), or $\tau \approx \tfrac14$ ($X$ model);
a somewhat similar 
behavior can also be seen for some (convergent) truncated expansions,
and seems to reflect the above sign alternation.

What happens when the underlying lattice is taken to be 2-dimensional?
The functional forms under investigation here [Eqs. (\ref{modelsdef})]
diverge to $+\infty$ for some mutual orientations, and, on the other 
hand, Refs. \cite{rMWlate,rMWarXiv} address the general case
of continuous functions of the scalar product and 
Ref. \cite{rMWlate} can even allow for some 
singularities;  as far as we could check, the divergent
behavior of the models under investigation here does not fit into the 
framework of weak singularity  conditions used
in section 2.2  of Ref. \cite{rMWlate}.
More explicitly, based on the  series expansion
in Eq. (\ref{vsexp}), one could try to realize a decomposition 
of $V(\tau)$ along the lines of Ref. \cite{rMWlate}, 
(sect. 2.2, around their Eqs. (24) to (26), page 441--443),  by
choosing a (large) positive integer $q$ and rewriting Eq. 
(\ref{Vmodel}) as
\begin{equation}
V(\tau) = \ln(2) + \mathcal{V}_q+r_q;
\label{eqdec01}
\end{equation}
the  divergent  term $r_q$ would then be positive around
$\Delta = \pi$, and its sign would {\it not} agree with
the hypotheses stipulated for theorem 1, singular case, in 
Ref. \cite{rMWlate}, where the  small singular term
in the interaction is  written (their notation) 
$$
-v(\phi),~v(\cdot) \ge 0.
$$
Thus there appears to be no available mathematical theorem 
entailing a Mermin-Wagner-type result in this case,
although it has been  conjectured (expectation is not 
calculation) that, in the thermodynamic limit, orientational order is 
also destroyed at all finite temperatures; (see. e.g. Ref. 13 in Ref. 
\cite{rsing4}); on the other hand, at least for the $V$ case,
one might expect a BKT behavior, since the singularity of the 
potential should ultimately strengthen short-range correlations.

\section{Simulation aspects and finite--size scaling theory}\label{comptaspect}
For $D=2$, the three models $V$, $W$
and $X$ [Eqs. (\ref{modelsdef})] were treated by simulation.
Calculations were carried out using periodic boundary conditions, and  
on samples consisting of $N=L^2$ particles, with
$L=40,60,80,100,120,160$.
Simulations, based on standard Metropolis updating algorithm, were
carried out in cascade, in order of increasing temperature $T$; equilibration
runs took between 25000 and 50000 cycles, where one cycle
corresponds to
$2N$ attempted Monte Carlo steps, including sublattice
sweeps (checkerboard decomposition
\cite{rmult01,rmult02,rmult03,rmult04}), and production runs 
took between 500000 and 1500000.

Subaverages for evaluating statistical
errors were calculated over macrosteps consisting of 
1000 cycles. Calculated quantities include 
the potential energy (in units $\epsilon$ per particle), and
derivative with respect to temperature 
based on the fluctuation formula
\begin{equation}
U^* = \frac{\langle H \rangle}{N},
\end{equation}
and
\begin{equation}
C^* = \frac1{NT^2}\left(\langle H^2\rangle - \langle H 
\rangle^2\right),
\end{equation} 
with
\begin{equation}
H  = \sum_{\left\{j < k \right\}} F(\tau_{jk}),
\end{equation}
where $\sum_{\left\{j < k \right\}}$ denotes sum over all distinct
nearest--neighbouring pairs of lattice sites.

As for orientational quantities, such as mean magnetization and corresponding 
susceptibilities \cite{rchi1,rchi2}, they can be expressed in general by
\begin{subequations}
\begin{equation}
\mathbf{P} = \sum_{k=1}^N \mathbf{w}_k,
\label{b21}
\end{equation}
\begin{equation}
M = \frac{1}{N} \langle |\mathbf{P}|  \rangle,
\end{equation}
\begin{equation}
M_2 = \frac{1}{N} \langle \mathbf{P} \cdot \mathbf{P} \rangle,~
\label{b22}
\end{equation}
\end{subequations}
\begin{subequations}
\begin{equation}
\chi_1 = \left\{ 
\begin{array}{ll}
\beta \left(M_2 - N M^2\right),&T < T_c \\[0.5cm]
\beta M_2,&T \ge T_c
\end{array}
\right.,
\label{b23}
\end{equation}
where  $\beta=1/T$,
and $T_c$ denotes  the critical temperature;
since $|\mathbf{P}| \le N$ [Eq. \eqref{b21}], we have
\begin{equation}
M_2  \le N \quad \mathrm{and} \quad ~\chi_1 \le \beta N.
\label{b24}
\end{equation}
\end{subequations}
Notice that Eq. (\ref{b23}) involves a true ordering transition temperature
$T_c$: in our case, for models $V$ and $W$, we found consistent evidence 
of the absence of orientational order at all finite temperatures (see also 
following Section), i.e. $T_c=0$, and selected the definition of $\chi_1$
accordingly.
Model $X$ [Eq. (\ref{Xmodel})], on the other hand, possesses even symmetry, and its 
second-- and fourth--rank order parameters
$\overline{P}_2$ and $\overline{P}_4$, as well as the corresponding susceptibility
$\chi_2$, were calculated as discussed in Ref. \cite{pre77051704};
notice that, in this case 
\begin{equation}
\chi_2 \le \beta N.
\label{b25}
\end{equation}
We also calculated various short--range order parameters, defined by
\begin{equation}\label{SOP}
\sigma_J= \langle \mathcal{E}_J(\tau_{jk}) \rangle,
\end{equation}
measuring correlations between corresponding pairs of unit vectors
associated with nearest--neighbouring sites; here
$\mathcal{E}_J(\tau)$ denote appropriate orthogonal polynomials
[see Eq. (\ref{eqOP}) in Appedix \ref{AppA}], and we 
chose $J=1,2$ for both $V$ and $W$ models, 
and $J=2,4$ for the $X$ model.

In the quest for the possible occurrence of a phase transition in the
models investigated here, we will analyse the
simulations data via the finite--size scaling (FSS) theory for
continuous phase transitions -- second order and BKT (infinite order) 
\cite{rmult04,newman_monte_1999,chamati2013}. According to
FSS hypothesis when a system is restricted to a
finite geometry (a square of area $L^2$ in the present
case) its thermodynamic quantities acquire a size
dependence with a behavior that is tightly related to the order of the
phase transition. It is worth mentioning that finite-size effects 
become important when the correlation length is of the same
order as the linear size of the system.
To be more specific we give details based on the behavior
of the susceptibility.

In the vicinity of a bulk critical point $T_c$ the (magnetic) 
susceptibility diverges
against the reduced temperature $t=1-\tfrac{T}{T_c} \ll 1$ according the
scaling law $\chi_1\sim|t|^{-\gamma}$ with the critical exponent 
$\gamma>0$. For a finite-size system it turns into
\begin{equation}\label{chiso}
\chi_1(L,T)=L^{\gamma/\nu}\Theta_\chi (tL^{1/\nu}),
\end{equation}
where $\nu$ measures the degree of divergence of the distance over 
 which the spins are correlated, \textit{i.e.} the correlation length 
$\xi\sim|t|^{-1/\nu}$ with $\nu>0$. The function $\Theta_\chi(x)$ is a universal
function depending on the gross features of the system, but not of its
microscopic details.

On the other hand, when a BKT transition takes place, the susceptibility 
of the bulk system diverges exponentially
\begin{equation}\label{chibkt}
\chi_{BKT}\sim a_\chi 
\exp\left[b_\chi\left(T-T_{BKT}\right)^{-\tfrac12}\right], 
\quad T_{BKT} \lesssim T
\end{equation} 
as we approach $T_{BKT}$ and is infinite in the BKT phase with a
quasi--long range order. For a finite 
system however the divergence is rounded and the susceptibility is 
finite [Eqs. \eqref{b24} and \eqref{b25}]. 
In the vicinity of the bulk BKT temperature the correlation length is proportional 
to the system's linear size and the susceptibility scales like
\begin{equation}\label{chifss}
\chi_{BKT}\sim L^{2-\eta_{BKT}(T)}.
\end{equation}
At the transition temperature $\eta_{BKT}=\tfrac14$.

Expressions \eqref{chiso} and \eqref{chifss} are valid asymptotically in
the vicinity of the transition temperature \textit{i.e.} when both the
sample size $L$ and the correlation length $\xi$ are very large, but
their ratio $\tfrac{\xi}L$
is finite. In this limit the universal scaling behavior is not
affected by the finite-size effects.

\section{Simulation results and FSS analysis} 
\label{results}
Simulation results obtained for the three investigated models
turned out to exhibit broad qualitative similarities, to be  contrasted
to their regular counterparts (see following discussion).
\subsection{The magnetic models $V$ and $W$} \label{results-mag}
Simulation results for various observables, obtained for the two models 
$V$ and $W$, were found to
exhibit a recognizable qualitative similarity over a wide temperature
range, so that, in some cases, only $V$ results will be presented in the
following.

Simulation data for the potential energies of both models 
(not shown here) were found to evolve with tempereture in a gradual, 
monotonic way, and to be essentially independent of sample sizes,
to within statistical errors falling below $0.1 \%$.

\begin{figure}[h!]
\resizebox{\columnwidth}{!}{\includegraphics{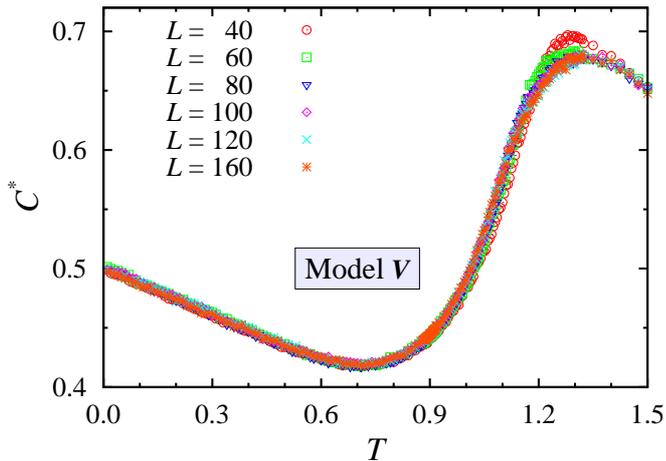}}
\caption{(Color online) The specific heat of model $V$ for different 
sample sizes against temperature;
statistical errors (not shown here) range between $1$ and $5 \%$.
Meaning of symbols: red circles: $L=40$;  green squares: $L=60$;
blue triangles: $L=80$; magenta diamonds: $L=100$: cyan crosses,
red asterisks: $L=160$.}
\label{cvd2n2}
\end{figure}

\begin{figure}[h!]
\resizebox{\columnwidth}{!}{\includegraphics{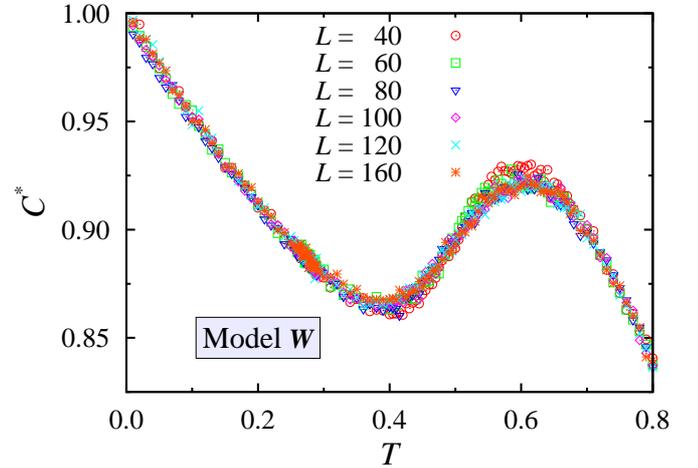}}
\caption{(Color online) The specific heat of model $W$ for different sample sizes against temperature;
statistical errors (not shown here) range between $1$ and $5 \%$;
same meaning of symbols as in FIG. \ref{cvd2n2}.}
\label{cvd2n3}
\end{figure}

As for the configurational specific heat $C^*$ (see FIG. \ref{cvd2n2}
for model $V$, and FIG. \ref{cvd2n3}, for model $W$), related to
thermal fluctuations of the potential energy, the plots showed that
$C^*$ starts with a maximum at $T=0$, and first decreases to a broad
minimum (say at $T^{\prime})$; it then increases to another maximum
(say at $T^{\prime \prime}$); here the associated statistical errors
range between $1$ and $5\%$, and results are only mildly affected by
sample size.  We found $T^{\prime} \approx 0.75,~T^{\prime \prime}
\approx 1.2$ for the $V$ model, and $T^{\prime} \approx 0.4,~T^{\prime
\prime} \approx 0.62$ for the $W$ counterpart; upon extrapolating the
low--temperature results to $T=0$, we estimate the corresponding
zero--temperature values to be $\tfrac12$ and $1$, respectively;
notice also that the zero--temperature value for the $W$ model (but
not for the $V$ model) corresponds to the global maximum; on the other
hand, $T^{\prime \prime}$ for the $V$ model (but not for the $W$
model) corresponds to the global maximum. The same behaviour
was found by estimating the specific heat via numerical
differentiation of the internal energy.

A finite-size analysis of the configurational specific heat according
to corresponding scaling behavior compatible with \eqref{chiso} ruled
out the existence of a second order phase transition in both models. A
similar analysis was performed on the magnetization and the
susceptibility for both models, but no scaling was achieved.

\begin{figure}[h!]
\resizebox{\columnwidth}{!}{\includegraphics{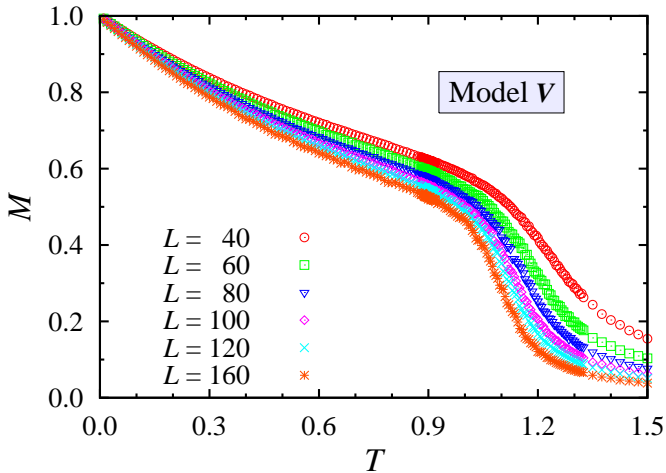}}
\caption{(Color online) Simulation results for the  magnetisation of model $V$ 
obtained with different sample sizes; same meaning of symbols as in
FIG. \ref{cvd2n2}.}
\label{magd2n2}
\end{figure}

Simulation results for the magnetization obtained with both models
(see e.g. FIG. \ref{magd2n2} for model $V$) showed a decreasing
behavior as a function of temperature for a given sample size;
at each examined nonzero temperature, they kept
decreasing with increasing sample size;
low--temperature results appear to extrapolate
to $M=1$ at $T=0$ for all examined sample size, as expected.

Low--temperature simulation results for $M$ and for both models $V$
and $W$ were found to exhibit a
power--law decay with increasing sample size; recall that
the spin--wave analysis worked out in Ref. \cite{tobochnik_spinwave_1979} 
for the regular counterpart [Eq. \eqref{reg01}] predicts the low--temperature result
\begin{equation}
M \approx \left(2L^2\right)^{-\tfrac T{8\pi}}.
\end{equation}
Our data at a given temperature were well fitted in a log-log scale by the relation
\begin{equation} \label{eqaddres02}
\ln M = -a \ln L + b, \qquad ~a>0,
\end{equation}
where the ratio $\tfrac{a(T)}T$ was found to increase with temperature,
and to become constant in the low--temperature limit.

\begin{figure}[h!]
\resizebox{\columnwidth}{!}{\includegraphics{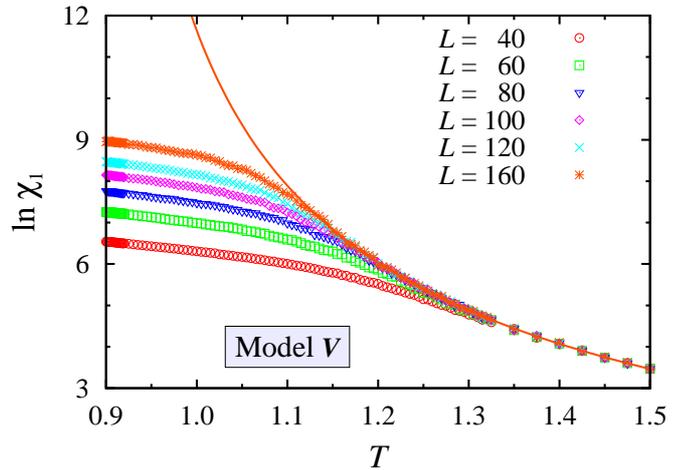}}
\caption{(Color online) Simulation results for the susceptibility $\chi_1$
of model $V$ obtained with different sample sizes; same
meaning of synbols as in FIG. \ref{cvd2n2}; 
Assuming a BKT transition and fitting the largest sample size $L=160$
(upper continuous orange curve)
to the bulk behavior of the susceptibility leads a transition at
$T_{BKT}=0.883\pm0.007$.}
\label{chi2d2n2}
\end{figure}

\begin{figure}[h!]
\resizebox{\columnwidth}{!}{\includegraphics{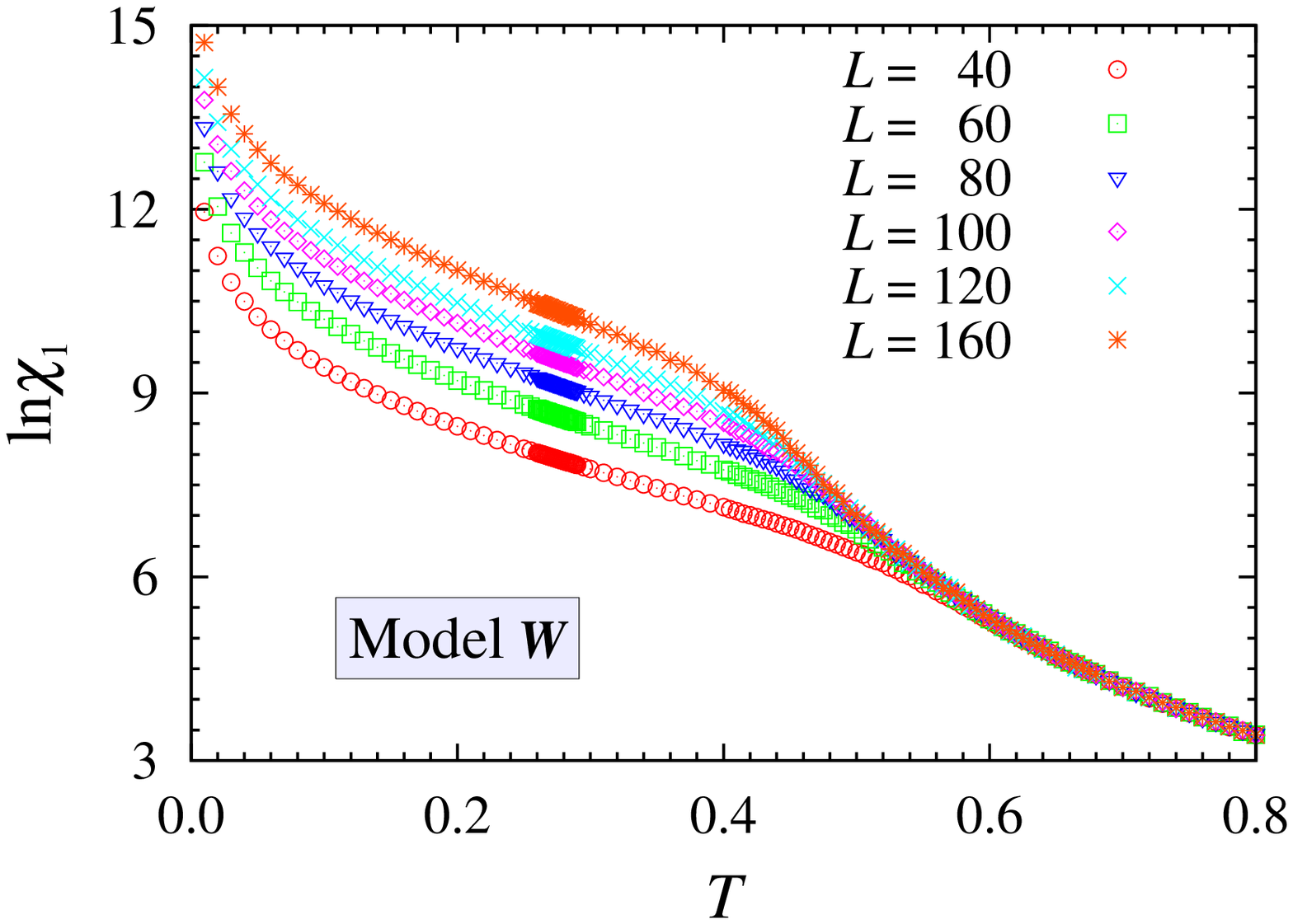}}
\caption{(Color online) Simulation results for the susceptibility $\chi_1$
of model $W$ obtained with different sample sizes; same
meaning of symbols as in FIG. \ref{cvd2n2}.
}
\label{chi2d2n3}
\end{figure}

The thermal fluctuations of the magnetization for both models $V$ and
$W$ i.e. their magnetic susceptibilities (actually $\ln\chi_1$) are
presented in FIGs. \ref{chi2d2n2} and \ref{chi2d2n3}. At low
temperatures the susceptibility keeps growing with sample size for
both models, within the constraint of \eqref{b24}, whereas at higher
temperatures it becomes independent of sample size;
the temperatures $T_{ch}$ where this change of scaling
behavior first becomes recognizable  are 
$T_{ch} \approx 1.3 > T^{\prime \prime}$ for model $V$, 
and $T_{ch} \approx 0.56 < T^{\prime \prime}$  for model $W$, respectively.

This specific behavior suggests a BKT transition from a quasi-long
range ordered phase at low temperatures to a disordered phase at
higher ones. Assuming such a transitional behavior,
 we have fitted the data of the
largest sample size ($L=160$) to expression \eqref{chibkt} for
the bulk susceptibility and found the results of Table \ref{table1},
as crude estimates (see also below).

\begin{table}[h!]
\caption{Estimates of the parameters in Eq. \eqref{chibkt}
obtained by fitting to data for the largest sample size for
models $V$, $W$ and $X$ assuming they exhibit a BKT transition.\label{table1}}
\begin{ruledtabular}
\begin{tabular}{cccc}
Model	&  $\ln(a_\chi)$  &    $b_\chi$   & $T_{BKT}$           \\
\hline
V	&  $-3.21\pm0.15$ & $5.29\pm0.15$ &  $0.873\pm0.007$    \\
W	&  $-3.93\pm0.14$ & $5.40\pm0.14$ &  $0.259\pm0.007$    \\
X	&  $-2.33\pm0.05$ & $3.03\pm0.04$ &  $0.347\pm0.003$    \\
\end{tabular}
\end{ruledtabular}
\end{table}

We analyzed the behavior of the susceptibility $\chi_1$ according to
the finite-size scaling ansatz \eqref{chifss} in the vicinity of
$T=0.9$ for model $V$ and of $T=0.28$ for model $W$; we first  carried
out a linear fit of $\ln\chi_1$ vs. $\ln L$ and estimated the critical
exponent $\eta$ from the slope of the curves corresponding to
different temperatures. The values obtained are presented in Tables
\ref{tablev} and \ref{tablew}, for models $V$ and $W$, respectively. A
nonlinear fit, based on Eq. \eqref{chifss} was performed as well, and
yielded results in agreement with these ones. Thus the transition
temperatures are most likely at $T_{BKT}= 0.910 \pm 0.005$ and
$T_{BKT}= 0.275 \pm 0.005$ for models $V$ and $W$, respectively. The
discrepancy between these values and those in Table \ref{table1}
points to the presence of huge finite-size effects: recall that Eq.
\eqref{chibkt} holds in the thermodynamic limit only, but was applied
here to the largest investigated sample size in the hope to gain
insights in the transitional behavior of the models considered here.

For the regular counterpart of model $V$ the
configurational specific heat was found to exhibit a sharp maximum at about
15\% \cite{tobochnik_spinwave_1979} above the BKT transition. In Refs.
\cite{chamati_2006,chamati_2007} we have investigated the impact of
diluted random impurities on the transition temperature. In Ref. \cite{chamati_2006}
we have found a broad peaks about 5\% above the BKT transition, and in
Ref. \cite{chamati_2007} we found a sharper one about 2\% above the
transition temperature. Here we find a maximum at about 40\% above
$T_{BKT}$. All these results show that the maximum of the specific
heat is always above the transition temperature.
As for $T_{ch}$, we could not find in the Literature any estimate
for the regular counterpart [Eq. (\ref{reg01})];
thus additional simulations were run
for the named regular model, carried out with
the same sample sizes as for the three 
singular models, and using
overrelaxation \cite{rov1,rov2,rov3,rov4,rov5};
the estimate $T_{ch} \approx 1.05$ was obtained. 

\begin{table}[h!]
\caption{Estimates of $\eta$ for model $V$ obtained via a log-log fit 
according to Eq. \eqref{chifss} for different temperatures along with 
the corresponding error $\delta\eta$.
\label{tablev}}
\begin{ruledtabular}
\begin{tabular}{ccccccccc}
$T$    & 0.890 & 0.895 & 0.900 & 0.905 & 0.910 & 0.915 & 0.920 \\
\hline
$\eta$ & 0.246 & 0.244 & 0.248 & 0.240 & 0.250 & 0.251 & 0.257 \\
$\delta\eta$
       & 0.007 & 0.006 & 0.006 & 0.006 & 0.005 & 0.004 & 0.006
\end{tabular}
\end{ruledtabular}
\end{table}
\begin{table}[h!]
\caption{Estimates of $\eta$ for model $W$ obtained via a log-log fit 
according to Eq. \eqref{chifss} for different temperatures along with 
the corresponding error $\delta\eta$.
\label{tablew}}
\begin{ruledtabular}
\begin{tabular}{cccccccc}
$T$    & 0.265 & 0.270 & 0.275 & 0.280 & 0.285 & 0.290 & 0.295 \\
\hline                                                               
$\eta$ & 0.239 & 0.243 & 0.249 & 0.257 & 0.271 & 0.273 & 0.285  \\
$\delta\eta$
       & 0.004 & 0.004 & 0.004 & 0.006 & 0.006 & 0.005 & 0.007  
\end{tabular}
\end{ruledtabular}
\end{table}

\begin{figure}[h!]
\resizebox{\columnwidth}{!}{\includegraphics{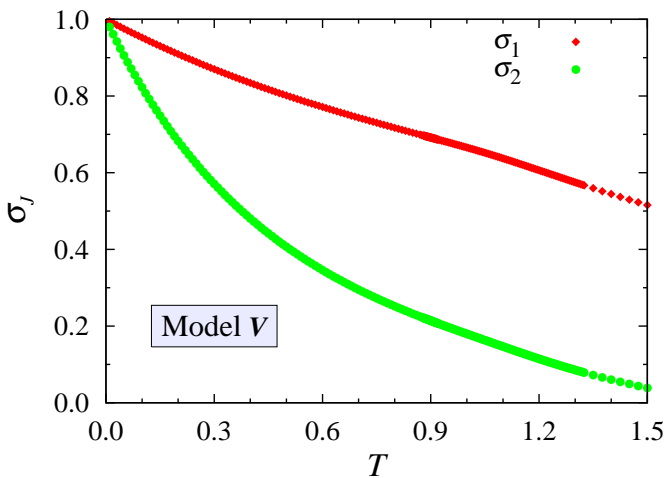}}
\caption{(Color online) Simulation results for the short-range order parameters
\eqref{SOP} of model $V$ obtained with the largest sample
size; meaning of symbols: red diamonds (upper curve) $\sigma_1$;
green circles (lower curve): $\sigma_2$.}
\label{sigmasd2n2}
\end{figure}

Simulation data for the short--range order parameters defined in 
\eqref{SOP} were found to be independent of sample size, and to 
decrease with temperature in a gradual and continuous way, paralleling 
the potential energy data; results obtained with the largest sample 
size of model $V$ are collected on FIG. \ref{sigmasd2n2}. 

\subsection{The two-dimensional nematic model $X$} \label{results-nem}
\begin{figure}[h!]
\resizebox{\columnwidth}{!}{\includegraphics{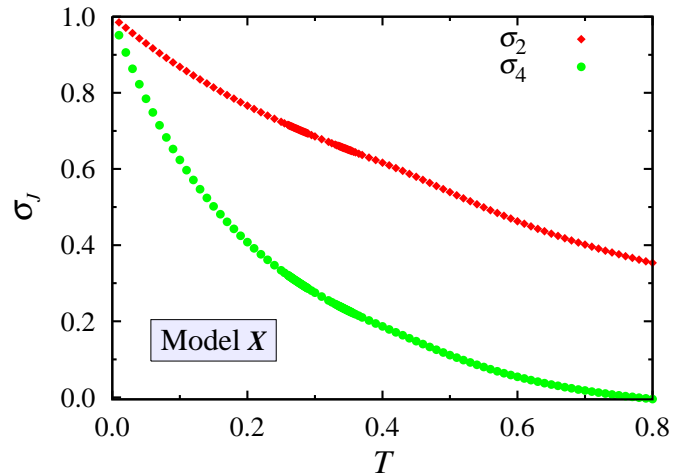}}
\caption{(Color online) Simulation results for the short-range order parameters
\eqref{SOP} of model $X$ obtained with the largest sample
size; meaning of symbols: red diamonds (upper curve) $\sigma_2$;
green circles (lower curve): $\sigma_4$.}
\label{sigmaslnabs}
\end{figure}

\begin{figure}[h!]
\resizebox{\columnwidth}{!}{\includegraphics{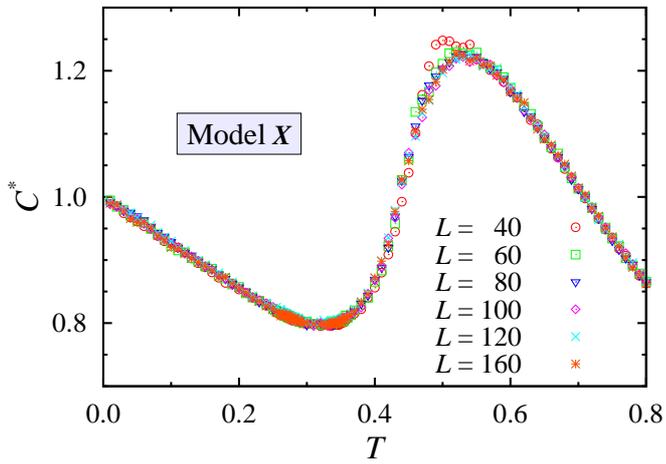}}
\caption{(Color online) The specific heat of model $X$ for different sample sizes 
against temperature;
statistical errors (not shown here) range between $1$ and $5 \%$;
same meaning of symbols as in FIG. \ref{cvd2n2}.}
\label{cvlnabs}
\end{figure}

Simulation results for the $X$ model were also found to exhibit a
remarkable qualitative similarity with the ones obtained for their
magnetic counterparts. Data for the potential energy (not shown) as
well as for the short--range order parameters (FIG. \ref{sigmaslnabs})
were found to be independent of sample size, and to evolve with
temperature in a gradual and monotonic way. The temperature dependence
of the specific heat corresponded to its magnetic counterpart (FIG.
\ref{cvlnabs}); here also the associated statistical errors were found
to range between $1$ and $5\%$, and the results appeared to be only
mildly affected by sample size. The plot started with the value $1$
at $T=0$, decreased with increasing temperature reaching a broad
minimum at $T^{\prime} \approx 0.3$, and then its global maximum at
$T^{\prime \prime} \approx 0.5$. it is  worth mentioning that a
quite similar behavior was obtained by numerical differentiation
of the potential energy. Notice also that, in the three cases,
sample--size effects on the results become more pronounced about
$T^{\prime \prime}$. Here we anticipate that neither the results for
the specific heat nor those corresponding to the second--rank order
parameter $\overline{P}_2$ or to the susceptibility $\chi_2$ could
obey the scaling behavior characteristic of a second order phase
transition.

\begin{figure}[h!]
\resizebox{\columnwidth}{!}{\includegraphics{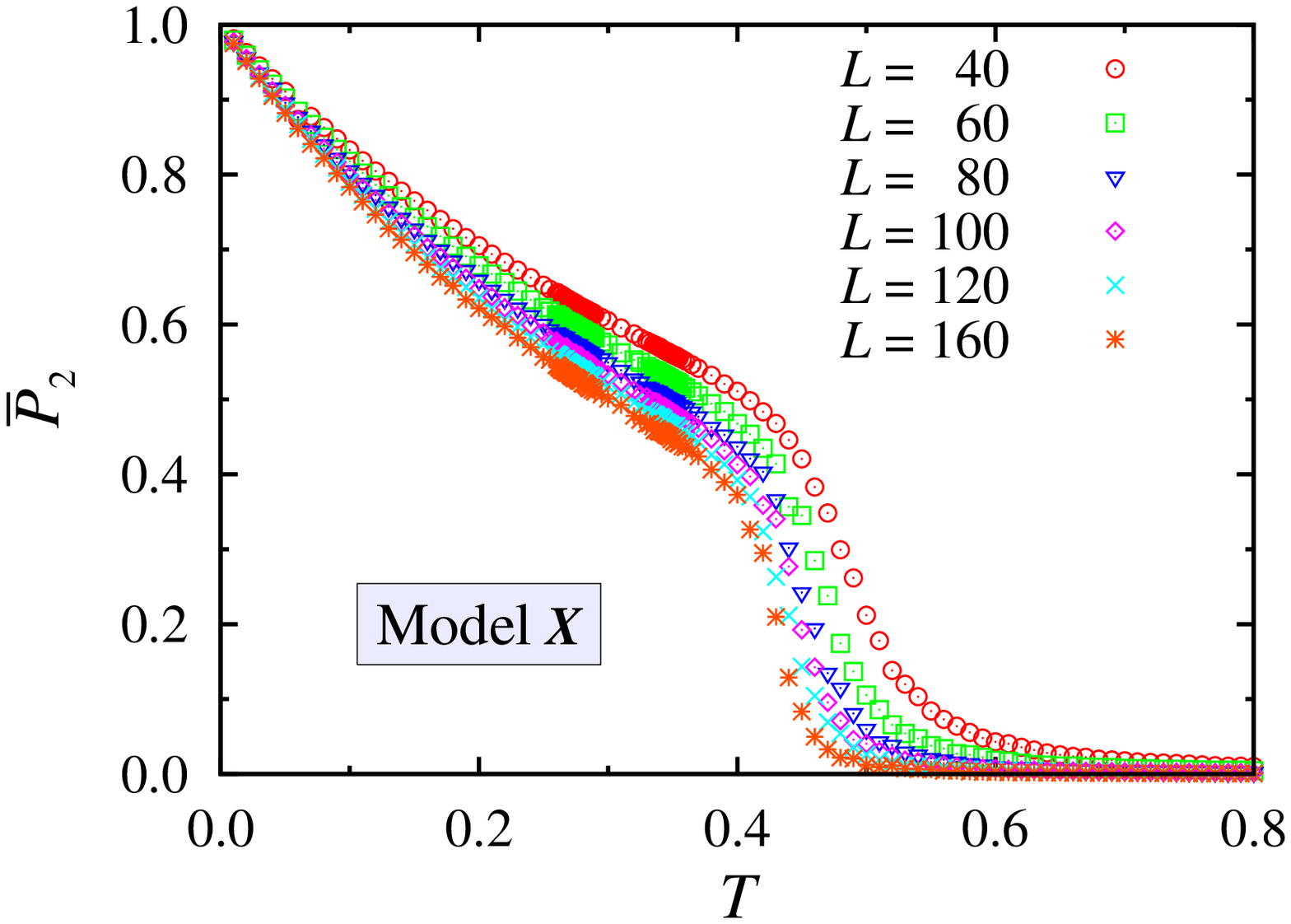}}
\caption{(Color online) Simulation results for the second--rank order parameter 
$\overline{P}_2$ of model $X$ obtained with different sample sizes;
same meaning of symbols as in FIG. \ref{cvd2n2}.} 
\label{p2lnabs}
\end{figure}

\begin{figure}[h!]
\resizebox{\columnwidth}{!}{\includegraphics{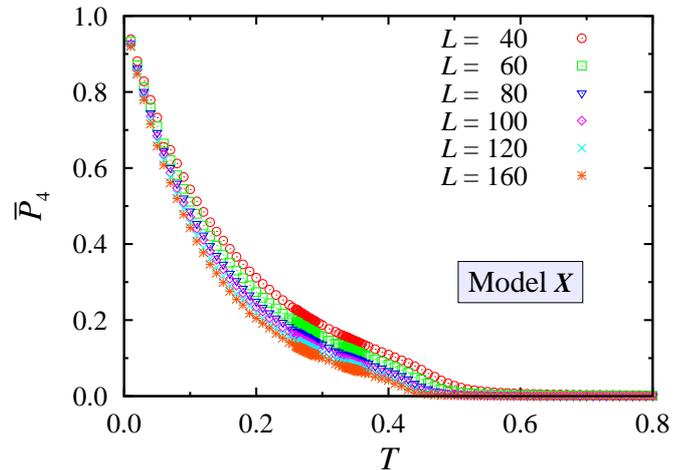}}
\caption{(Color online) Simulation results for the fourth--rank order parameter 
$\overline{P}_4$ of model $X$ obtained with different sample sizes;
same meaning of symbols as in FIG. \ref{cvd2n2}.} 
\label{p4lnabs}
\end{figure}

Simulation results for the order parameters $\overline{P}_J$, ($J=2,4$)
were also found to decrease with increasing temperature for each 
sample size, and to decrease with increasing sample size at each 
nonzero temperature (FIG. \ref{p2lnabs} and FIG. \ref{p4lnabs}).
At all investigated temperatures the results for the nematic order
parameters $\overline{P}_M$, ($M=2,4$) exhibited a power--law decay with
increasing sample size. At a given temperature these were well fitted 
to the corresponding relations
\begin{equation} \label{eqaddres03}
\ln\overline{P}_J = -b_{J1} \ln L + b_{J0}, \qquad ~b_{J1}>0.
\end{equation}
The coefficients $b_{J1}(T)$ were found to increase with $T$,
and to become proportional to $T$ to within statistical
errors in the low temperature region.
The results obtained from Eq. (\ref{eqaddres03}) show that both order
parameters vanish in the thermodynamic limit i.e. $L\to\infty$;
such a behavior is in agreement with the spin wave theory for magnetic 
systems discussed above.


\begin{figure}[h!]
\resizebox{\columnwidth}{!}{\includegraphics{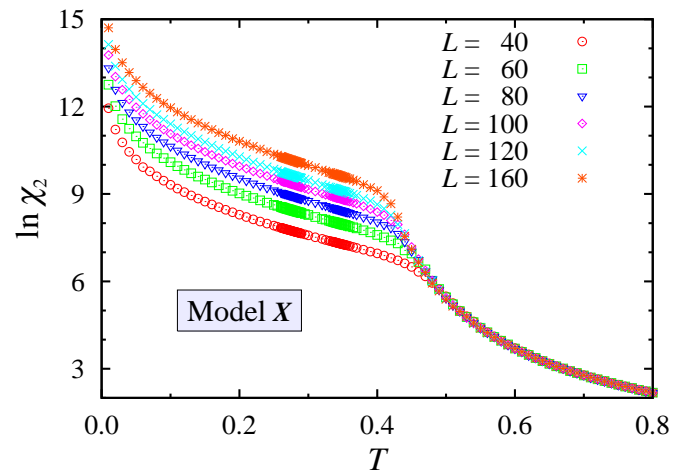}}
\caption{(Color online) Simulation results for the susceptibility $\chi_2$
of model $X$ obtained with different sample sizes; same meaning of symbols
as in FIG. \ref{cvd2n2}.}
\label{chi2lnabs}
\end{figure}

Simulation results for $\ln \chi_2$ versus $T$ (FIG.
\ref{chi2lnabs}) showed a low-temperature regime where they kept
increasing with increasing sample size, and then
became independent of sample size at higher temperatures;
the temperature $T_{ch}$ where  this change of scaling first becomes
recognizable was $T_{ch} \approx 0.45 < T^{\prime \prime}$;
this behavior also parallels the one observed for the two magnetic
counterparts.

By fitting the data obtained at high temperatures for our largest
sample size ($L=160$) to
expression \eqref{chibkt} of the susceptibility, we obtain the results
reported in Table \ref{table1} with a transition temperature
$T_{BKT}=0.347 \pm 0.003$.

Upon applying the finite--size--scaling analysis with data for all
sample sizes to the susceptibility given by Eq. \eqref{chifss}, we end
up with the results of Table \ref{tablex} with an estimate of the
transition temperature $\Theta_{BKT} = 0.275 \pm 0.005$ for model X.
Here again we observe a discrepancy between the result obtained by
fitting the bulk expression of the susceptibility to the data for the
largest size and the FSS analysis. This may be traced back to the huge
finite-size effects. 

\begin{table}[h!]
\caption{Estimates of $\eta$ for Model $X$ obtained via a log-log fit according to
Eq. \eqref{chifss} for different temperatures along with the
corresponding error $\delta\eta$.
\label{tablex}}
\begin{ruledtabular}
\begin{tabular}{cccccccc}
$T$    & 0.260 & 0.265 & 0.270 & 0.275 & 0.280 & 0.285 & 0.290 \\
\hline
$\eta$ & 0.237 & 0.243 & 0.248 & 0.250 & 0.258 & 0.266 & 0.269 \\
$\delta\eta$
       & 0.006 & 0.005 & 0.005 & 0.006 & 0.004 & 0.006 & 0.003
\end{tabular}
\end{ruledtabular}
\end{table}

\subsection{Comparisons with the regular counterparts} \label{comparisons}
As for the regular counterparts [Eqs. \eqref{regall}],
the existence of a BKT transition
is by now a well-known result for planar rotators [Eq. \eqref{reg01}], and
an estimate of the transition temperature to be found
in the Literature is $T_{BKT} = 0.8929 \pm 0.0001$
\cite{hasenbusch_binder_2008,arisue_2009}; $T_{BKT}$ found
for the $V$ model is about $2 \%$ higher than the corresponding
value for the regular counterpart. 

On the other hand,
available evidence does not seem to support a BKT scenario
for the classical $O(3)$ Heisenberg regular counterpart [Eq.
\eqref{reg02}]. Various authors (see, e.g., Ref.
\cite{butera1990})
have argued that the model does not exhibit such a transition;
the opposite view has been put forward
by Patrascioiu and Seiler, in a series of papers (see e.g.
\cite{patras2002,*Patrascioiu1991173}); examples of the
resulting debate can be found in or via  Refs.
\cite{PhysRevE.90.032109}.

The nematic case [Eq. \eqref{reg03}] has been studied for some 30
years
\cite{Solomon1981492,Sinclair1982173,Fukugita1982209,Chiccoli1988298,kogan1990,khveshchenko1991,kunz1989,*Kunz1991299,*PhysRevB.46.662,PhysRevB.46.11141,Caracciolo1993815,Mukhopadhyay1997,bulgadaev2001,Mondal2003397},
and a BKT scenario has been proposed
by various Authors: a recent estimate of the transition temperature is
$T_{BKT}=0.548 \pm 0.002$ \cite{Mondal2003397}, with the $C^*$ maximum
at $T^{\prime \prime} \approx 0.57$, and $T_{ch} \approx T^{\prime
\prime}$;
on the other hand, some other Authors claim that the named model [Eq.
\eqref{reg03}] does not exhibit any critical transition, but its
low--temperature behavior is rather characterized by a crossover from
a disordered phase to an ordered phase at zero temperature
\cite{PhysRevE.78.051706,farinas-sanchez_critical_2010}.

\begin{table}[h!]
\caption{A summary of characteristic temperatures for the three models examined
by simulation in the present work; see text for definitions.}
\label{temps}
\begin{ruledtabular}
\begin{tabular}{clllc}
Model	&  $T^{\prime}$  &  $T^{\prime \prime}$ & $T_{ch}$   & $T_{BKT}$     
\\
\hline
V & $\approx 0.75$ & $\approx 1.25$ & $\approx 1.3$ & $0.910 \pm 0.005$ 
\\
W   &   $\approx 0.4$   &   $\approx 0.62$ & $\approx 0.56$   & $0.275 \pm 0.005$
\\
X & $\approx 0.3$ &  $\approx 0.5$ & $ \approx 0.45$ & $0.275 \pm 0.005$  
\\
\end{tabular}
\end{ruledtabular}
\end{table}

These comparisons (see also Table \ref{temps}) suggest
that, on the one hand, the singular character of the interaction
may bring about a BKT behavior where the regular counterpart
does not support it (W model); on the other hand, the effect on $T_{BKT}$
appears to be milder where
the regular counterparts already support
this transitional behavior, and this we interpret as a reflection
of the potential features 
pointed out previously (Sect. \ref{models}), in the discussion of
Eqs. (\ref{modelsexp}) and of FIGS. \ref{modelv}, \ref{modelw} and
\ref{modelx}.

In contrast to the regular counterparts, where the temperature
dependence of $C^*$ shows a simple maximum, upon
increasing temperature from $T=0$, the three singular models investigated
here exhibit first a minimum and then a maximum of $C^*$; 
this behavior also appears connected with the potential features
discussed in Sect. \ref{models}.

\section{Summary and conclusions}\label{summary}
We have revisited and generalized a previously studied model
\cite{rsing1,rsing2} and defined a few others,
whose pairwise interactions
are isotropic in spin space and restricted to nearest neighbours;
in contrast to other extensively studied models, their
functional forms contains logarithmic singularities which, so to speak,
do not disturb the thermodynamics. When $D=1$, the above models could be
solved in closed form, in terms of
Gamma, Beta and Polygamma functions, and
were found to produce orientational disorder 
and no phase transition,  at all finite temperatures, 
in the thermodynamic limit.
Some of the above models have been studied by simulation for $D=2$:
among a few candidates (see Section \ref{models}), we had chosen 
those functional forms which  strongly favour mutual parallel
orientations, thus strengthening (at least) short--range
correlations; in the absence of more stringent rigorous results, 
the obtained simulation results point to orientational disorder at all
finite temperatures, and suggest a BKT scenario in the three cases;
we hope to carry out a more thorough  simulation study
of the models.

Moreover, the
investigated models contain logarithmic singularities,
causing them to slowly diverge as $\tau \rightarrow -1$ or
$\tau \rightarrow 0$; on the other hand, 
comparison with the regular counterparts and with the
above constrained models (Section \ref{models})
leads one to speculate as to
what happens if the interaction potential is chosen to be  more
confining, \textit{i.e.} made more rapidly divergent  as $\tau$
moves away from $\tau_{min}$
(actually, a multitude of  such functional forms
can be envisaged); preliminary
work along these lines has been started, and its results will be reported
in due course.

\acknowledgments

The present extensive calculations were carried out, on,
among other machines,
workstations, belonging to the Sezione di Pavia of
Istituto Nazionale di Fisica Nucleare (INFN); allocations of computer
time by the Computer Centre of Pavia University and CILEA
(Consorzio Interuniversitario Lombardo
per l'Elaborazione Automatica, Segrate - Milan),
as well as by CINECA
(Centro Interuniversitario Nord-Est di Calcolo Automatico,
Casalecchio di Reno - Bologna),
and CASPUR (Consorzio interuniversitario per le 
Applicazioni di Supercalcolo per Universit\`a e Ricerca, Rome)
are gratefully acknowledged.
This work was supported by the exchange program between Bulgaria \&
Germany (DNTS/Germany/01/2).

\appendix

\section{Exact solutions for $D=1$} \label{AppA}
Some available exact results in one dimension are recalled here;
when $D=1$ (hence $\mathbf{x}_j\equiv j \in \mathbb{Z}$),  
for a linear sample consisting of $N$ spins, the Hamiltonian reads
\begin{equation}
H = \sum_{j=1}^N F\left(\mathbf{w}_j \cdot \mathbf{w}_{j+1}\right),
\label{eq02}
\end{equation}
where we assume periodic boundary conditions i.e.
$\mathbf{w}_{N+1}=\mathbf{w}_1$;
the corresponding overall partition functions can be calculated
exactly, and this is usually realized based on the underlying $O(n)$
symmetry, by means of an appropriate coordinate transformation
(i.e., geometrically, by taking each spin $\mathbf{w}_j$ as defining the 
reference axis for the next one $\mathbf{w}_{j+1}$) 
\cite{rexa1,rexa2,rexa3,rexa4,rexa5,rexa6};
the corresponding overall partition function reduces 
to the $N-$th power [or $(N-1)-$th power if one uses free boundary
conditions] of a single-particle
quantity, to be denoted here by $q(T)$; in formulae
\begin{subequations}
\label{eqs03}
\begin{eqnarray}
q(T)&=&\frac{1}{2\pi} p(T),
\label{eqs03a}
\\
p(T)& = & \int_0^{2 \pi} \exp(-\beta F(\cos s)) ds, \quad n=2,
\label{eqs03b}
\end{eqnarray}
\end{subequations}
and
\begin{subequations}
\label{eqs04}
\begin{eqnarray}
q(T)&=&\frac{1}{2} p(T),
\label{eqs04a}
\\
p(T)&  = &\int_{-1}^{+1} \exp(-\beta F(s)) ds, \quad n=3,
\label{eqs04b}
\end{eqnarray}
\end{subequations}
where $\beta =1/T$; correlation functions are defined by
\begin{equation}
G_J(m) = \langle \mathcal{E}_J(\mathbf{w}_j \cdot \mathbf{w}_k )\rangle,~
\mathrm{as~function~of}~m=|\mathbf{x}_j-\mathbf{x}_k|;
\label{eqcorr}
\end{equation}
here $J$ is
a strictly positive integer, and $\mathcal{E}_J(\tau)$ denote
appropriate orthogonal polynomials, i.e.
\begin{equation}
\mathcal{E}_J(\tau)=
\left\{
\begin{array}{ll}
T_J(\tau) = \cos( J \arccos(\tau)),&~n=2 \\[0.35cm]
P_J(\tau),&~n=3
\end{array}
\right. ;
\label{eqOP}
\end{equation}
here $T_J(\ldots)$ denote Chebyshev polynomials of the first kind, and
$P_J(\ldots)$ denote Legendre polynomials. For general $D$, and
when $F(\tau)$ is not an even function of its argument,
the simplest correlation function is $G_1(r)$;
for $D=1$, the definition in Eq. (\ref{eqcorr}) simplify to
\begin{equation}
G_J(m) = \langle \mathcal{E}_J(\mathbf{w}_j \cdot \mathbf{w}_k) \rangle,~
\mathrm{as~function~of}~m=|j-k|;
\label{eq04}
\end{equation}
and $G_1(m)$ reduces  to the $m-$th power of the quantity
\begin{equation}
c_1(T) = \frac{r_1(T)}{p(T)},
\end{equation}
where
\begin{subequations}
\begin{eqnarray}
r_1(T) & = & \int_0^{2 \pi} \cos s \exp(-\beta F(\cos s))  ds,~ n=2,
\label{eqs05a}
\\
r_1(T) & = & \int_{-1}^{+1} s \exp(-\beta F(s)) ds, \quad n=3.
\label{eqs05b}
\end{eqnarray}
\end{subequations}
The corresponding susceptibility is given by \cite{rchi1,rchi2}
[see also the following Eqs. (\ref{b22}) and (\ref{b23})]
\begin{eqnarray}
\chi_1 &=& \frac{\beta}{N} \left\langle \sum_{j=1}^N \sum_{k=1}^N 
(\mathbf{w}_j \cdot \mathbf{w}_k) \right\rangle \nonumber \\
&=& \frac{\beta}{N} \sum_{j=1}^N \sum_{k=1}^N G_1(|j-k|) \nonumber \\
&=& \frac{\beta}{N} \sum_{j=1}^N \sum_{k=1}^N c_1 ^{|j-k|};
\label{eqchia}
\end{eqnarray}
hence, in the large--$N$ limit,
\begin{equation}
\chi_1 = \beta \frac{1+c_1}{1-c_1}.
\label{eqchi1b}
\end{equation}
These quantities have been calculated 
in the Literature in a few cases, where
$F(\tau)$ is a simple polynomial of its argument. i.e.
$F=\pm \tau~(n=2,3),~F=\pm P_2(\tau)~(n=3)$
\cite{rexa1,rexa2,rexa3,rexa4,rexa5,rexa6,rexa7};
in the latter cases $F(\tau)$ is an even function of its argument,
so that the simplest relevant correlation function is
\begin{equation}
G_2(m) = \langle P_2(\mathbf{w}_j \cdot \mathbf{w}_k) \rangle,~
\mathrm{as~function~of}~m=|j-k|,
\label{eq05a}
\end{equation}
which similarly reduces to the $m-$th power of
\begin{subequations}
\begin{eqnarray}
c_2(T) & = & \frac{r_2(T)}{p(T)},
\label{eq05b}
\\
r_2(T) & = & \int_{-1}^{+1} P_2(s) \exp(-\beta F(s))  ds,\quad n=3;
\nonumber\\
\label{eq05c}
\end{eqnarray}
\end{subequations}
in the large--$N$ limit, the corresponding susceptibility reads
\begin{equation}
\chi_2 = \beta \frac{1+c_2}{1-c_2}.
\label{eqchi2}
\end{equation}

Notice that the  continuity of $F(\tau)$ implies 
convergence and regularity of $q(T)$; moreover
the definitions entail $|c_1(T)|<1$ or $|c_2(T)|<1$ 
at all finite temperatures;
thus leading to the well known results related to the
absence of phase transitions at all finite temperatures,
orientational disorder in the thermodynamic limit 
at all finite temperatures, and exponential decay with
distance for the absolute value of the  correlation functions;  
actually,
these results may  also hold under weaker conditions on $F(\tau)$.

There also exist in the literature a few lattice--spin models
involving mild integrable singularities, i.e.
defined by bounded and generally continuous functions of
the scalar products,
which still allow usage of the method  outlined here when $D=1$;
one such case is the sign or
step model \cite{rstep1,rstep2,rstep3,rstep4,rstep5,rstep6,rstep7}, defined by
\begin{equation}
F(\tau) = \pm\ \mathrm{sign}(\tau);
\label{eqsign}
\end{equation}
the model was solved exactly for $D=1$  and $n \ge 2$
\cite{rstep3}, and proven to remain orientationally disordered
even at $T=0$, where calculations
in Ref. \cite{rstep3} yield for the ferromagnetic case
\begin{equation}
G_1(1) = \frac{1}{\sqrt \pi} \, \frac{\Gamma(n/2)}{\Gamma((n+1)/2)};
\end{equation}
for $D=n=2$ there is consistent
evidence of orientational disorder at all temperatures, as well as of 
the existence of a BKT
transition \cite{rstep5,rstep6,rstep7}.

We notice in passing that other extensions of Eq. \eqref{eqsign}
can be envisaged, e. g.
\begin{equation}
F(\tau)= \pm \mathrm{sign}(P_J(\tau)),~n=3,
\label{extendsign}
\end{equation}
where, say, $J=2,3,4$; when $D=1$, the resulting 
partition functions can be worked out in closed form as well.

The effect of divergences in $F(\tau)$ was seldom  
investigated, and we shall be 
considering here some extensions of Eqs. (\ref{Vmodel}) and (\ref{Wmodel}), 
in addition to Eq. (\ref{Xmodel}),
\begin{subequations}
\label{mypots}
\begin{eqnarray}
V_I(\tau)& =& -\ln(1+I \tau), \quad n=2,
\label{mypot-V}
\\
W_I(\tau)& =& -\ln(1+I\tau), \quad n=3,
\label{mypot-W}
\end{eqnarray}
\end{subequations}
where $I=\pm 1$ defines the ferro- or antiferro-magnetic character of the
interaction.
Both $V_I(\tau)$ and $W_I(\tau)$ attain their minimum when $\tau=I$,
and slowly diverge to $+\infty$ as $\tau \rightarrow -I$;
$X(\tau)$ attains its minima when $\tau=\pm 1$
and slowly diverges to $+\infty$ as $\tau \rightarrow 0$;
the above functions are bounded from below, continuous almost everywhere,
and possess integrable singularities; moreover,
their functional forms turn out to be computationally convenient
for $D=1$. 
Two other related models can be defined as well,
by combinining ferro-- and antiferro--magnetic cases of $V_I(\tau)$ with equal
positive weights, and similarly for $W_I(\tau)$; in formulae:
\begin{subequations}
\begin{eqnarray}
A_2(\tau)& =& -\ln(2(1-\tau^2)), \quad n=2,
\label{mypot-A2}
\\
A_3(\tau)& =& -\ln(1-\tau^2), \quad n=3.
\label{mypot-A3}
\end{eqnarray}
\end{subequations}
Both $A_2(\tau)$ and $A_3(\tau)$ are even functions of their
argument, attaining their minimum for $\tau=0$ and diverging to
$+\infty$ for $|\tau| \rightarrow +1$; the letter $A$ in the names recalls
their antinematic character.
Actually, further generalizations of the $V_I$ models are possible, i.e.
\begin{equation}
V_{I,K}(\tau) = -\ln(1+I \cos(K \Delta_{jk})), \quad n=2
\label{mypot-VG}
\end{equation}
where $K$ is an arbitrary, strictly positive, integer, and
$V_{I,1}=V_{I}$.
By now it has been known for some time
that interaction models only differing in the value of $K$
produce the same partition functions, and that the resulting
orientational properties can be defined in a way independent of $K$
\cite{genBKT,rmap1,rmap2}; for more details see  Appendix \ref{AppB}.
A few specific cases are listed here
\begin{subequations}
\label{eqIM}
\begin{eqnarray}
V_{+1,1}(\tau)&=&-\ln(1+\cos(\Delta_{jk})),
\label{eqIM01}
\\
V_{-1,1}(\tau)&=&-\ln(1-\cos(\Delta_{jk})),
\label{eqIM02}
\\
V_{+1,2}(\tau)&=&-\ln(1+ \cos(2 \Delta_{jk})),
\label{eqIM03}
\\
V_{-1,2}(\tau)&=&-\ln(1-\cos(2 \Delta_{jk})).
\label{eqIM04}
\end{eqnarray}
The standard trigonometric identity
$$
\cos(2x) = 2\cos^2 x-1
$$
entails that
\begin{eqnarray}
V_{+1,2}(\tau)&=&-\ln( 2 \cos^2\Delta_{jk}),
\label{eqIM05} \\
V_{-1,2}(\tau)&=&-\ln(2 \sin^2 \Delta_{jk});
\label{eqIM06}
\end{eqnarray}
\end{subequations}
one recognizes that $V_{+1,2}$ defines the $2-$component counterpart
of the $X$ model, and that $V_{-1,2}$ essentially coincides with
$A_2$.
 
The above models can be solved explicitly, as worked out
in the following: notice also that some qualitative results can be obtained
in a more direct and elementary way, e.g., for $V_{I,1}(\tau)$,
\begin{subequations}
\begin{eqnarray}
p(T) &=&\int_0^{2\pi} (1+I \cos s)^{\beta} ds  \nonumber\\
&=& \int_{-\pi/2}^{+\pi/2}
\left[(1+\cos s)^{\beta}+(1-\cos s)^{\beta}\right] ds;~ \nonumber \\
\end{eqnarray}
and, for the correlation function, 
\begin{eqnarray}
r_1(T)  &=&  \int_0^{2\pi} \cos s  (1+I \cos s)^{\beta} ds \nonumber\\
&=& I \int_{-\pi/2}^{+\pi/2}
\cos s \left[ (1+\cos s)^{\beta}-(1-\cos s)^{\beta} \right]
ds;\nonumber\\
\end{eqnarray}
\end{subequations}
since $|\cos s| \le 1$, the above equations entail $|c_1(T)|<1$ at all finite 
temperatures.
A similar approach can be used  $W_I(\tau)$, i.e.
\begin{subequations}
\begin{eqnarray}
p(T) &=&\int_{-1}^{+1} (1+I s)^{\beta} ds  \nonumber\\
&=&  \int_{0}^{+1}
\left[(1+s)^{\beta}+(1-s)^{\beta}\right] ds;~
\end{eqnarray}
and, for the correlation function, 
\begin{eqnarray}
r_1(T)  &=&  \int_{-1}^{+1} s  (1+I s)^{\beta} ds \nonumber\\
&=& I \int_{0}^{+1}
s \left[ (1+s)^{\beta}-(1-s)^{\beta} \right] ds;
\end{eqnarray}
\end{subequations}
since $|s| \le 1$, the above equations entail $|c_1(T)|<1$ at all 
finite temperatures.

Notice that, for each of the two functional forms
(\ref{mypot-V}) or (\ref{mypot-W}), and
in the absence of an external field, the two possible
choices for I define  models producing the same partition functions
and correlation functions related by appropriate numerical factors
(equivalent by spin--flip symmetry).

The above models can be solved explicitly in terms of known special
functions with well defined analytic properties, and some of them yield 
results involving the functions: Gamma
$$
\Gamma(z) =  \int_0^{+\infty} s^{z-1} \exp(-s) ds,
$$
Beta
$$
B(x,y) = B(y,x) = \frac{\Gamma(x) \Gamma(y)}{\Gamma(x+y)};
$$
and Polygamma
$$
\Psi(l,z) = \frac{d^{l+1}}{d z^{l+1}} \ln \Gamma(z).
$$
Here $x,~y,~z$ are complex variables with $\Re(x)>0,~\Re(y)>0,\Re(z)>0$,
and $l$ denotes a nonnegative integer 
\cite{rAS,rGR}; let us also recall
that $\Gamma\left(\frac{1}{2}\right)=\sqrt{\pi}$.

The above properties of $V$ models read 
\begin{subequations}
\label{vmod}
\begin{eqnarray}
p(T)&  = & 2^{\beta} \int_0^{2\pi} (\cos^2s)^{\beta} ds \nonumber\\
& = & 2^{\beta} \int_0^{2 \pi} (\sin^2 s)^{\beta} ds 
\nonumber
\\
~ & = & 2 \sqrt{\pi} 2^{\beta} \frac{\Gamma(\beta + \tfrac12)}{\Gamma(\beta + 1)}
\label{vmod-p}
\\
q(T) & = & \frac{\sqrt{\pi}}{\pi} 2^{\beta} \frac{\Gamma(\beta +
\tfrac12)}{\Gamma(\beta + 1)};
\label{vmod-q}
\end{eqnarray}
the configurational specific heat (in units $k_B$ per particle)
can be obtained via the appropriate 
derivatives of the partition function and reads
\begin{equation}
C^*  =  \frac1{T^2}\left[\Psi \left( 1,\frac1T+\frac12 \right)-
\Psi \left( 1,\frac1T \right)\right]+1
\label{vmod-c}
\end{equation}
For $V_{+1,2}$
\begin{equation}
c_2(T) =  \frac{\beta}{\beta + 1},
\end{equation}
and in general for $V_{I,K}$
\begin{equation}
c_K(T) =  I \frac{\beta}{\beta + 1},
\end{equation}
\end{subequations}
notice that $c_2$ for  $V_{+1,2}$ is the same as $c_1$ for $V_{+1,1}$.

The corresponding results for $W_I(\tau)$ are
\begin{subequations}
\label{wmod}
\begin{eqnarray}
q(T)  & = & \frac{2^{\beta}}{\beta +1},
\label{wmod-q}
\\
C^* & = &   \frac{1}{1+T^2},
\label{wmod-c}
\\
c_1(T)& = & I \frac{\beta}{\beta +2}.
\label{wmod-chi}
\end{eqnarray}
\end{subequations}

For $X(\tau)$ one finds 
\begin{subequations}
\label{xmod}
\begin{eqnarray}
p(T) = \frac{2}{\beta+1},
\label{xmod-q}
\\
C^* =  \frac{1}{1+T^2},
\label{xmod-c}
\end{eqnarray}
$X(\tau)$ is an even function of its argument, and the previous Eqs. 
(\ref{eq05b}) and (\ref{eq05c}) specialize to
\begin{eqnarray}
	c_2(T) & = & \frac{r_2(T)}{p(T)},
\\
r_2(T) & = & \int_{-1}^{+1} P_2(s) |s|^\beta ds,
\end{eqnarray} 
and eventually
\begin{eqnarray}
q(T)  & =  & \frac{1}{\beta+1},
\\
c_2(T)& = & \frac{\beta}{\beta +3 }.
\label{xmod-chi}
\end{eqnarray}
\end{subequations}
Notice also that  both $W_I(\tau)$ and $X(\tau)$ yield
the same expression for the configurational contribution to the specific
heat per particle [Eqs. \eqref{wmod-c} and \eqref{xmod-c}],
and produce rather similar expressions for $c_1$ [Eq. \eqref{wmod-chi}]
and $c_2$ [Eq. \eqref{xmod-chi}], respectively.
As for the  four $V$ models in Eqs. \eqref{eqIM}, let us recall that
models with the same $I$ and different $K$ produce the same partition
functions, and their orientational
properties can be defined
in a way independent of $K$, i.e. $G_m(r)$ for $V_{I,1}$
is the same as $G_{2m}(r)$ for $V_{I,2}$ 
\cite{genBKT,rmap1,rmap2}; on the other hand,
the above calculations also show that $V_{+1,1}$ and $V_{-1,1}$
produce the same partition functions and correlation functions
connected by appropriate sign factors; thus the four named
interaction models [Eqs. \eqref{eqIM}] produce one and the same partition 
function, and essentially the same orientational properties.

The corresponding properties for $A_3$ model can be obtained
in closed form as well;
\begin{subequations}
\label{a3mod}
\begin{equation}
q(T) = \frac{\sqrt{\pi}}{2 \pi} \frac{\Gamma(\beta
+1)}{\Gamma(\beta + \tfrac32)}
\label{a3mod-q}
\end{equation}
\begin{equation}
C^* =
\frac {1}{{T}^{2}} \left[ \Psi \left( 1,1+\frac1T\right)
-\Psi \left( 1,\frac32+\frac 1T \right) \right]
\label{a3mod-c}
\end{equation}
\begin{equation}
c_2(T)  = - \frac{\beta}{2 \beta + 3}.
\label{a3mod-chi}
\end{equation}
\end{subequations}
Notice that one can combine the potential models $X$ and $A_3$ to define
\begin{equation}
Y(\tau) =-\ln[\tau^2(1-\tau^2)],\quad ~n=3;
\label{eqY}
\end{equation}
in this case the interaction diverges to $+\infty$ when $\tau=0$
and $|\tau|=1$; on the other hand, by standard trigonometric
identities, one can recognize that the $n=2$ counterpart
corresponds to $V_{-1,4}$ within numerical factors.
The partition function of model $Y$ is
\begin{subequations}
\label{ymod}
\begin{equation}
q(T) = \frac{\sqrt{\pi}}{2} 4^{-\beta} \frac{\Gamma(2 \beta +1)}
{\Gamma(2 \beta+\tfrac32)},
\label{ymod-q}
\end{equation}
and the corresponding quantities are given by
\begin{equation}
C^*  =
\frac4{T^2} \left[\Psi \left( 1,1+\frac2T\right)
-\Psi \left( 1,\frac32+\frac2T \right) \right]
\label{ymod-cv}
\end{equation}
\begin{equation}
c_2(T) = \frac{1}{4+3T}.
\label{ymod-chi}
\end{equation}
\end{subequations}
In all of the above cases, $C^*$ was found to be a monotonic
decreasing function of temperature, in contrast to the regular
counterparts Eqs. (\ref{reg01}) and (\ref{reg03}), which 
produce a maximum of $C^*(T)$; on the other hand, Eq. (\ref{reg02})
also produces a monotonic decreasing behavior for $C^*(T)$.

In the main text we are simply referring to $V_{+1,1}$ as $V$ model,
and to $W_{+1}$ as $W$ model.
For $D=1$, the named models produce no phase transition
and no orientational order at finite temperatures in the thermodynamic 
limit; actually, some non--integrable 
singularities in $F(\tau)$ can produce the same qualitative
behavior as well; this  happens, for example, with
constrained models, defined as follows:
let $s_0$ denote a real number, $0 < s_0 < \pi,~\tau_0=\cos s_0$,
and let \cite{rMWlate,rPS,rMA,rWB}
\begin{equation}
F(\tau) = \left\{ 
\begin{array}{l l}
f(\tau) &,~+1 \ge \tau > \tau_0 \\[0.3cm]
+\infty &,~-1 \le \tau < \tau_0
\end{array}
\right .,
\label{eqclick}
\end{equation}
where $f(\tau)$ denotes some regular function of its argument
(see also below); in other words, the absolute value of the 
angle between the two interacting unit vectors,
defined modulo $2\pi$, is constrained to remain below the
threshold $s_0$.
Upon following the previous line of thought and applying 
Eqs. (\ref{eqs03}) to (\ref{eqs05b}), one can recognize that,
when $D=1$, functional forms like
Eq. (\ref{eqclick}) also  produce no phase transition
and no orientational order at finite temperatures in the thermodynamic
limit.
Models defined by Eq. (\ref{eqclick}) and  $D=n=2$ have also been
addressed:
for $f(\tau)=-\tau$, it was proven that,
when $s_0$ is sufficiently small, the correlation function 
$G_1(r)$ never decays exponentially with distance,
but obeys an inverse--square lower bound at all temperatures
\cite{rMWlate,rPS,rMA}; on the other hand, 
when $f(\tau)=0$ \cite{rWB}, the system is athermal, and there is a
simulation evidence of a BKT transition with $s_0$ as control parameter.

\section{Mapping between potential models} \label{AppB}
Consider the integral
\begin{equation}
\psi = \int_{0}^{2 \pi} \Phi(\cos s, \sin s)ds,
\end{equation}
where $\Phi$ denotes a sufficiently regular function, and let
\begin{equation}
\Psi_K = \int_0^{2\pi} \Phi(\cos K s,\sin Ks) ds,
\end{equation}
where $K$ is an arbitrary non--zero integer, and $\Psi_1=\psi$;
one can immediately verify that
\begin{equation}
\forall K \in \mathbb{Z} \setminus \{{0}\}, \Psi_K=\psi;
\end{equation}
consider now 
\begin{equation}
\Xi = \int_0^{2\pi} \exp( \pm i  \mu s) \Phi(\cos Ks, \sin Ks) ds,
\label{eqapp02}
\end{equation}
where $\mu>1$ denotes  an arbitrary positive integer,
and recall the identity
\begin{equation}
\sum_{j=1}^{j=\mu}\exp\left(\pm 2\pi i \frac{j}{\mu}\right) = 0. \quad \mu>1.
\end{equation}
Thus the value of $\Xi$ in Eq. (\ref{eqapp02}) is zero when
$\mu$ is not an integer multiple of $K$; 
on the other hand, when $\mu$ is an integer multiple of $K$, 
say $\mu =\lambda K$,
the value of $\Xi$ is again independent of $K$

\end{document}